\documentclass{cimento}

\usepackage{graphicx}  

\title{Out-of-equilibrium density dynamics of a spinful Luttinger liquid}
\author{Filippo Maria Gambetta\from{1}\from{2} \atque Sergio Porta\from{1}\from{2}}
\instlist{\inst{1} Dipartimento di Fisica, Universit\`a di Genova, Via Dodecaneso 33, I-16146 Genova, Italy.
		\inst{2}SPIN-CNR, Via Dodecaneso 33, I-16146 Genova, Italy. }
\PACSes{\PACSit{71.10.Pm}{Luttinger liquid}
\PACSit{73.21.La}{Electron states and collective excitations in quantun dots}
\PACSit{67.85.Lm}{Ultracold degenerate Fermi gases}
\PACSit{05.70.Ln}{Nonequilibrium Thermodynamics}}

\begin{document}

\maketitle

\begin{abstract}
Using a Luttinger liquid theory we investigate the time evolution of the particle density of a one-dimensional spinful fermionic system with open boundaries and subject to a finite-duration quench of the  inter-particle interaction. Taking into account also the turning on of an umklapp perturbation to the system Hamiltonian as a result of the quench, we study the possible formation of a Wigner molecule inside the system, focusing in particular on the sudden and adiabatic regimes. We show that the creation of this correlated state is essentially due to the propagation of ``light-cone'' perturbations through system which arise after both switching on and switching off the quenching protocol and that its behavior strongly depends on the quench duration.
\end{abstract}

\section{Introduction}
During the last decade the remarkable experimental improvements in the field of ultracold atoms renewed the interest in studying non-equilibrium dynamics of isolated quantum many-body systems~\cite{Bloch:2008,Polkovnikov:2011,expqq}. Among the latter, one-dimensional (1D) systems are an ideal playground to explore the interplay between non-equilibrium and interaction effects. Indeed, from the theoretical point of view, the study of out-of-equilibrium 1D systems can benefit from the generalization of tools developed to investigate their equilibrium properties to the non-equilibrium regime. In particular, the Luttinger liquid (LL) model~\cite{giamarchi}, which describes the equilibrium low energy sector of any 1D gapless system including the main effects of inter-particle interactions, proved to be very powerful also in exploring out-of-equilibrium properties. In particular, it allowed for a better understanding of many general features of non-equilibrium dynamics of isolated quantum systems~\cite{cazalilla, qqreview}. For instance, sudden interaction quenches in LLs provided the first analytical results about the issue of thermalization in quantum integrable systems~\cite{cazalilla, qqreview,gge} and showed very neat signatures of the ``light-cone'' (LC) effect in correlations functions~\cite{qqreview,lightcone1,Calabrese:2006}. Furthermore they allowed to determine how transport and spectral properties are modified far from equilibrium ~\cite{qqtransport,qqtransportnoi} and to investigate the crossover from sudden to adiabatic quenches~\cite{dora,Bernier:2014,Chudzinski:2016,Porta:2016}. Finally, the possibility of addressing also open boundary conditions~\cite{Fabrizio:1995} enabled to study effects of quantum quench in finite-size 1D systems~\cite{Porta:2016,Kennes:2013}. In the latter case, when a 1D fermionic system is confined in a small region of space, marked oscillations appear in the particle density. In particular, {\it Friedel} oscillations, due to reflections at the system edges, exist regardless of the interactions between fermions and give rise to $ N/D $ peaks in the particle density (with $ N $ the number of fermions in the system, assumed even for simplicity, and $ D $ the single-particle levels degeneracy)~\cite{Fabrizio:1995,Giuliani:2005}. On the other hand inter-particle interactions, for example by means of a small umklapp term in the Hamiltonian of system, could lead to the formation of a Wigner molecule~\cite{wigner,GindikinSablikov,Soffing:2009,wignernoi,Gambetta:2014} -- the finite-size counterpart of the Wigner crystal~\cite{Giuliani:2005,Wigner:1934,Schulz:1993} -- characterized by $ N $ peaks in the particle density, regardless of $ D $.
In a previous work~\cite{Porta:2016} the authors studied the density dynamics of a spinless LL after a quantum quench of the inter-particle interaction with finite time duration showing that it exhibits a sharp LC behavior. In this system $ D=1 $ and thus Friedel and Wigner oscillations possess the same number of peaks~\cite{Gambetta:2014}. In this paper we generalize our previous results to the spinful case (with $ D=2 $), where a competition between $ N/2 $ peaks of Friedel oscillations and $ N $ peaks of Wigner ones is expected in the particle density, focusing on both a sudden and an adiabatic turning on of a repulsive inter-particle interaction. Since a possible umklapp term in the Hamiltonian -- which in the equilibrium is responsible for the emergence of Wigner oscillations -- strongly depends on the strength of inter-particle interactions~\cite{Giuliani:2005,Soffing:2009,Schulz:1993}, the quench should also affects its structure and the associated Wigner contribution to the density. Consequently, we also consider a simultaneous quench of the umklapp term in the system Hamiltonian obtaining an overall enhancement of the LC effect already found in the spinless case~\cite{Porta:2016}. Furthermore, we show that the formation of a Wigner molecule is conveyed by the propagation of LC perturbations through the system and is deeply affected by the time duration of the quench. In particular, in the sudden case two LC perturbations emerge  from the boundaries of the system immediately after the quench and make the particle density oscillate periodically in time between the non-interacting profile, with Friedel oscillations only, and the correlated one, with a Wigner molecule clearly visible inside the system. This periodic behavior is due to the finite size of the system, which results in a finite recurrence time and implies that the latter will never reach a steady state~\cite{Bocchieri:1957}. Thus, in order to create a stable Wigner molecule, one should resort to a slower quenching protocol. In this case -- and for strong enough inter-particle interactions -- two LC perturbations emerge at the beginning of the quench and, after each subsequent passage, contribute to broaden and to split every initial Friedel peak into two well separated Wigner peaks. After a long enough quench a Wigner molecule has thus stabilized inside the system while non-equilibrium effects survives as two different LC perturbations that make the particle density weakly oscillate above the Wigner molecule configuration: The first ones are the continuation of the LC perturbations emerged at the beginning of the quenching protocol while the second ones arise from the boundaries when the quench stops.

The paper is organized as follows. In Sec.~\ref{sec:model} we introduce our model, describe the quenching protocol and the various contributions to the particle density. In Sec.~\ref{sec:results} we analyze the quench dynamics of the particle density after both a sudden and an adiabatic quench, focusing on the possible formation of a Wigner molecule inside the system. Finally, Sec.~\ref{sec:conclusions} contains our conclusions. 

\section{The Model} \label{sec:model}
We start from a non-interacting spinful LL with $ N_s $ fermions with spin $ s=\{\uparrow,\downarrow\}= \{+,-\} $ confined in the region $ 0\leq x \leq L $ and subject to open boundary conditions $ \psi_s(0)=\psi_s(L)=0 $, where $ \psi_s(x) $ is the single-particle wavefunction. For $ t<0 $ the system is described by the initial Hamiltonian (hereinafter $ \hbar=1 $)~\cite{giamarchi,Fabrizio:1995}
\begin{equation}
\label{eq:Hi}
H_i=\sum_{\nu={\rho,\sigma}}\left[\sum_{q>0}qv_F b^{\dagger}_{\nu,q}b_{\nu,q}+\mathcal{E}(N_\nu)\right],
\end{equation}
with momenta $ q=\pi n/L $ (with $ n $ a positive integer). Here, $ b_{\nu,q}=b_{\uparrow,q}+\epsilon_\nu b_{\downarrow,q} $ are the bosonic annihilation operators of charge ($ \nu=\rho $, $ \epsilon_\rho=1 $) and spin ($ \nu=\sigma $, $ \epsilon_\sigma=-1 $) degrees of freedom and $ \mathcal{E}(N_\nu)=\frac{\pi v_F}{4L}N_\nu^2 $ is the zero-mode energy , where $ N_\nu=N_\uparrow+\epsilon_\nu N_\downarrow $ are the number operators of charge and spin excitations. 

Starting from $ t=0 $ a repulsive interaction between the particles is turned on and brought to the final value $ g_\nu=v_F(K_\nu^{-2}-1)/2 $ by means of a general quenching protocol $ Q(t) $ with finite time duration $ \tau $ and such that $ Q(t\leq0)=0 $ and $ Q(t\geq\tau)=1 $. Here, $ K_\nu=(1+2g_\nu/ v_F)^{-1/2} $ is the final Luttinger parameter, which describes the strength of inter-particle interactions (with $ K_\nu<0 $ for repulsive interactions and $ K_\nu=1 $ for non-interacting particles)~\cite{giamarchi,Kleimann:2002}. 
The time-dependent Hamiltonian of the system is
\begin{equation}
H(t)=\sum_{\nu=\rho,\sigma}\left\{\sum_{q>0}q\left[v_\nu(t)b^{\dagger}_{\nu,q}b_{\nu,q}+\frac{1}{2}g_\nu(t)\left(b^{\dagger}_{\nu,q}b^{\dagger}_{\nu,q}+b_{\nu,q}b_{\nu,q}\right)\right]+\mathcal{E}(N_\nu)\right\},\label{eq:Ht}
\end{equation}
with $ v_\nu(t)=v_F+ g_\nu Q(t) $ and $ g_\nu(t)=-g_\nu Q(t) $. For any $ 0\leq t\leq\tau $ the Hamiltonian $ H(t) $ can be diagonalized by a time-dependent Bogolubov transformation to the basis of bosonic operators $ \bar{b}_{\nu,q;t} $~\cite{Bernier:2014,Porta:2016} obtaining
\begin{equation}
\label{eq:vinstgen}
H(t)=\sum_{\nu=\rho,\sigma}\left\{\sum_q q\bar{v}_\nu (t) \bar{b}^\dagger_{\nu,q;t}\bar{b}_{\nu,q;t}+\mathcal{E}(N_\nu)\right\},
\end{equation}
where $ \bar{v}_\nu(t)=v_F/\bar{K}_\nu(t) $ and $ \bar{K}_\nu(t)=[1+2g_\nu Q(t)/v_F]^{-1/2} $ are the {\em instantaneous} LL bosonic modes velocity and Luttinger parameter, respectively. Note that for $ t\geq\tau $ one has $ \bar{K}_\nu(t)=K_\nu $ and $ \bar{v}_\nu(t)=v_F/K_\nu\equiv v_\nu $.
The time evolution of bosonic operators $ b_{\nu,q} $ for $ t>0 $ can be obtained from the Heisenberg equation of motion~\cite{dora} and is given by
\begin{equation}
\label{eq:betaevolution}
b_{\nu,q}(t)=f_\nu(q,t)b_{\nu,q}+h_\nu^*(q,t)b_{\nu,q}^{\dagger},
\end{equation}
with the bosonic operators on the right hand side evaluated at $ t=0 $ and the functions $ f_\nu(q,t) $ and $ h_\nu(q,t) $ satisfying the relation $ |f_\nu(q,t)|^2-|h_\nu(q,t)|^2=1\ \forall q,t $. For the sake of simplicity, in the following we will consider a linear ramp quench with~\cite{dora,Porta:2016}
\begin{equation}
\label{eq:protocol}
Q(t) = \left\{\begin{array}{ll}
0 &\mathrm{for}\ t<0\ \mathrm{(region\ I)},\\
t/\tau &\mathrm{for}\ 0\leq t\leq\tau\ \mathrm{(region\ II)},\\
1 &\mathrm{for}\ t>\tau\ \mathrm{(region\ III)},
\end{array}\right.
\end{equation}
and, in particular, we will focus on the sudden and the adiabatic quench regimes~\cite{Porta:2016}. The former is obtained in the limit $ \tau\rightarrow0 $~\cite{cazalilla, qqreview} while the latter is defined by the condition $ \delta_\nu\gg L/\pi $~\cite{Porta:2016}, where we have introduced the parameter $  \delta_\nu=\frac{2}{3}v_F\tau(K_\nu^{-2}-1)^{-1} $.
In both these regimes the functions $ f_\nu(q,t) $ and $ h_\nu(q,t) $, which encode the entire dynamics of the bosonic operators $ b_{\nu,q} $, can be obtained by a straightforward generalization of the spinless case and we thus refer the interested reader to Ref.~\cite{Porta:2016} for further details. 

\subsection{Bosonization and density operator}\label{subsec:bosonization}
 \noindent Following the standard bosonization prescriptions~\cite{giamarchi}, we decompose the fermionic field $ \Psi_s(x) $ in right-- [$ R_s(x) $] and left--  [$ L_s(x) $] moving fields. Working in the Heisenberg picture, we thus write $  \Psi_s(x,t)=R_s(x,t)+L_s(x,t) $. The open boundary conditions $ \Psi_s(0,t)=\Psi_s(L,t)  $, $ \forall t $, imply that $ L_s(x,t)=-R_s(-x,t) $~\cite{Fabrizio:1995}, with the bosonized right--moving field given by~\cite{giamarchi, Fabrizio:1995}
\begin{equation}
 \label{eq:bosR}
 R_s(x,t)=\frac{F_s(t)}{\sqrt{2\pi\alpha}}e^{i \pi N_s x/L}e^{i \phi_s(x,t)}.
 \end{equation}
Here, $ \alpha $ is a small length cutoff, $ \phi_s(x,t)=[\phi_\rho(x,t)+s\phi_\sigma(x,t)]/\sqrt{2} $, where $ \phi_\nu(x,t)=\sum_{q>0}\sqrt{\pi/L q}\ e^{-\alpha q/2}[e^{iqx}b_{\nu,q}(t)+e^{-iqx}b^\dagger_{\nu,q}(t)] $
are the charge and spin bosonic fields, and $ F_s(t) $ are the time-evolved Klein factors~\cite{giamarchi}. The particle density operator  is~\cite{giamarchi,Fabrizio:1995}
 \begin{equation}
 \rho(x,t)=\rho_{\mathrm{LW}}(x,t)+\rho_{\mathrm{F}}(x,t),  \label{eq:rho}
 \end{equation}
 with
 \begin{eqnletter}
 \label{eq:rhoin}
 \rho_{\mathrm{LW}}(x,t) &=&\frac{N_{\rho}}{L}-\frac{\sqrt{2}}{\pi}\partial_x\Phi_\rho^a(x,t),\label{eq:rhoLW}\\
 \rho_{\mathrm{F}}(x,t) &=&-\sum_{s}\frac{N_s}{L}\cos\left[\mathcal{L}_1(N_s,x)-2\Phi_s^a(x,t)\right],\label{eq:rhoFriedel}
 \end{eqnletter}
 long-wave and Friedel contributions, respectively. Here, $  \mathcal{L}_j(N,x)=\frac{2\pi N x}{L}-jf(x) $, $  \Phi_{\bullet}^a(x,t)=[\phi_{\bullet}(-x,t)-\phi_{\bullet}(x,t)]/2 $ and $
 f(x)=\arctan\{\sin(2\pi x /L)/[e^{\alpha \pi/L}-\cos(2\pi x/L)]\}$.
 Working in the zero-temperature limit and assuming the system prepared in the ground state of $ H_i $ for $ t<0 $, one can evaluate the quantum average of Eq.~(\ref{eq:rho}) making use of Eq.~(\ref{eq:betaevolution}) and following the standard bosonization procedure~\cite{giamarchi, Porta:2016,Fabrizio:1995,GindikinSablikov}. The long-wave term simply evaluates to $ N_\rho/L $ while the Friedel one is given by
 \begin{equation}
 \langle\rho_{\mathrm{F}}(x,t)\rangle_i=-\sum_{s}\frac{N_s}{L}[E_\rho(x,t)E_\sigma(x,t)]^{1/2}\cos\left[\mathcal{L}_1(N_s,x)\right],\label{eq:rhoF}
 \end{equation}
 where $ \langle...\rangle_i  $ represents the quantum average over the ground state of $ H_i $ and we set $ \alpha=L/(\pi N_s) $~\cite{GindikinSablikov}.
 The time dependence is entirely contained in the {\em envelope function}~\cite{Porta:2016}
 \begin{equation}
 \label{eq:envelope}
 E_\nu(x,t)=\exp\left[-\frac{2\pi}{L}\sum_{q>0}\frac{e^{-\alpha q}}{q}\sin^2(qx)|D_\nu(q,t)|^2\right],
 \end{equation}
with $ D_\nu(q,t)=f_{\nu}(q,t)-h_{\nu}(q,t) $, which for $ t>\tau $ is a periodic function with period $ \mathcal{T}_{\nu}=L/v_\nu $ due the finite size of the system~\cite{Bocchieri:1957}. 
 \noindent We recall that for a standard non-quenched LL with Luttinger parameter $ K $ one has $ E^{\mathrm{nq}}_{\nu}(x,t)\equiv E^{\mathrm{nq}}(x) $, with~\cite{Fabrizio:1995, Gambetta:2014}
 \begin{equation}
 \label{eq:Estd}
 E^{\mathrm{nq}}(x)=\exp\left[-\frac{2\pi}{L}\sum_{q>0}\frac{e^{-\alpha q}}{q}\sin^2(qx)\right]=\frac{\sinh(\pi\alpha/2L)}{\sqrt{\sinh^2(\pi\alpha/2L)+\sin^2(\pi x/L)}}.
 \end{equation}
With a straightforward generalization of the spinless case procedure~\cite{Porta:2016} one obtains the following expressions for the envelope function of Eq.~(\ref{eq:envelope}) in the sudden quench regime,
\begin{equation}
E^{\mathrm{sq}}_{\nu}(x,t)=[E^{\mathrm{nq}}(x)]^{(1+K_\nu^{2})/2}\left[\frac{E^{\mathrm{nq}}(v_\nu t)}{\sqrt{E^{\mathrm{nq}}(x-v_\nu t)E^{\mathrm{nq}}(x+v_\nu t)}}\right]^{(K_\nu^{2}-1)/2},\label{eq:envsq}
\end{equation} 
and in the two regions -- defined in Eq.~(\ref{eq:protocol})-- of the adiabatic one,
\begin{eqnletter}
E^{\mathrm{II,ad}}_{\nu}(x,t)&=&[E^{\mathrm{nq}}(x)]^{\bar{K}_\nu(t)}\left[1-\frac{\bar{K}_\nu(t)}{12\pi\delta_\nu }\mathcal{C}(x,\ell_\nu(t))\right],\label{eq:envadII}\\
E^{\mathrm{III,ad}}_{\nu}(x,t)&=&[E^{\mathrm{nq}}(x)]^{K_\nu}\left[1-\frac{K_\nu}{12\pi\delta_\nu}\mathcal{C}(x,v_\nu(t-\tau)+\ell_\nu(\tau))\right.\label{eq:envadIII}\\
&+&\left.\frac{K_\nu^{4}}{12\pi\delta_\nu}\mathcal{C}(x,v_\nu(t-\tau))\right].\nonumber
\end{eqnletter} 
Here,  $ \ell_\nu(t)=\delta_\nu[\bar{K}_\nu^{-3}(t)-1]  $, $ \mathcal{C}(x,y)=2\mathcal{D}(y)-\mathcal{D}(y+x)-\mathcal{D}(y-x)$, with $ \mathcal{D}(y)=\mathrm{Im}[\mathrm{Li}_2(e^{i2\pi y-\pi \alpha })] $ and $ \mathrm{Li}_2(x) $ the dilogarithm function (see Appendix of Ref.~\cite{Porta:2016} for further details).
 \subsection{Inclusion of the Wigner contribution}
In an interacting spinful LL with open boundary conditions the presence of an umklapp term in the Hamiltonian could lead to the formation of a Wigner molecule which manifests as density oscillations with a wave vector $ 2\pi N_\rho /L $~\cite{wigner,Soffing:2009,wignernoi,Schulz:1993}. Since this term strongly depends on the inter-particle interaction, we also address the impact of the quantum quench on it. For the sake of simplicity, in the following we will consider a quench in the charge sector only\footnote{Note that the Wigner contribution is not affected by a quench in the spin sector and thus what follows remains valid also in the general case.}, i.e. with $ K_\sigma=1$, and we will assume for the umklapp perturbation the same quenching protocol $ Q(t) $ [see Eq.~(\ref{eq:protocol})] as the inter-particle interaction one. Generalizing the analysis of Ref.~\cite{Soffing:2009}, we add to the Hamiltonian of Eq.~(\ref{eq:Ht}) a small umklapp perturbation,
 \begin{eqnarray}
H_U(t)&=&-\frac{g_U Q(t)}{2} \int_{0}^{L}dx\,\left[e^{i 2 N_\rho x/L}L_{\downarrow}^\dagger(x)R_{\downarrow}(x)L^\dagger_{\uparrow}(x)R_{\uparrow}(x)+\mathrm{H.c.}\right]\\
&=&-\frac{g_UQ(t)}{(2\pi\alpha)^2}\int_{0}^{L}dx\,\cos\left[\mathcal{L}_2(N_\rho,x)-2\sqrt{2}\Phi_\rho^a(x)\right],\nonumber
 \end{eqnarray}
with $\Phi_{\rho}^a(x,t)$ defined below Eq.~(\ref{eq:rhoin}). Working in the interaction picture with respect to the umklapp perturbation, the time evolution of the pre-quench ground state, $ |{\bf 0}_i\rangle $, is given by 
$|{\bf 0}^I_i(t)\rangle =|{\bf 0}_i\rangle +i g_U \sum_{{\bf n}_i}\int_{0}^{t}dt'\,\langle{\bf n}_i|H_{U}^I(t')|{\bf 0}_i\rangle$,
where $ |{\bf n}_i\rangle $ is a generic excited state of the unperturbed Hamiltonian. The correction $ \langle\delta \rho_{\mathrm{LW}}(x,t)\rangle_i=\langle {\bf 0}^I_i(t)| \rho_{\mathrm{LW}}(x,t)|{\bf 0}^I_i(t)\rangle-\langle \rho_{\mathrm{LW}}(x,t)\rangle_i $ to the average long-wave density of Eq.~(\ref{eq:rhoLW}) due to the presence of the umklapp perturbation evaluates to
\begin{eqnarray}
\langle\delta \rho_{\mathrm{LW}}(x,t)\rangle_i
&\approx&i\frac{4g_U}{L(2\pi\alpha)^2}\int_{0}^{x}dy\int_{0}^{t}dt'Q(t')\sin[\mathcal{L}(N_\rho,y)]\mathcal{F}_q(x,t;y,t')+\mathrm{H.c.}\label{eq:deltarho}\\
&\equiv&\lambda_{\mathrm{W}}\langle\rho_{\mathrm{W}}(x,t)\rangle_i, \nonumber
\end{eqnarray}
with $\mathcal{F}_q(x,t;y,t')=[E_\rho(y,t')]^{2}\sum_{q>0}e^{-\alpha q}\cos(qx)\sin(qy)D_\rho(q,t)D^*_\rho(q,t')$.
In the last line of the above equation we identified the perturbative correction with the Wigner contribution to the particle density -- which should be added to the quantum average of the density operator in Eq.~(\ref{eq:rho}) -- up to the multiplicative factor $ \lambda_{\mathrm{W}} $ [see Eqs.~(\ref{eq:rhoWsq}, \ref{eq:rhoWadII}, \ref{eq:rhoWadIII}) below for details]. Since the latter is proportional to $ g_U $ its exact value cannot be obtained in the framework of the LL theory and should be considered as a free parameter of the model~\cite{Soffing:2009,wignernoi,Schulz:1993}. In order to preserve the normalization of the total average density we also introduce an additional parameter $ \lambda_{\mathrm{F}} $, which measure the weight of Friedel contribution. Imposing the condition $ \langle\rho(0,t)\rangle_i=\langle\rho(L,t)\rangle_i=0 $ we obtain\footnote{Note that for $ t=0 $ one has $ \langle\rho_{\mathrm{W}}(x,t)\rangle_i=0$ and thus $  \langle\rho(x,t)\rangle_i=N_\rho/L+\langle\rho_{\mathrm{F}}(x,t)\rangle_i $, as expected. }
\begin{equation}
\langle\rho(x,t)\rangle_i=\frac{N_\rho}{L}+\left[1+\lambda_{\mathrm{W}}\frac{L}{N_\rho}\langle\rho_{\mathrm{W}}(0,t)\rangle_i\right]\langle\rho_{\mathrm{F}}(x,t)\rangle_i+\lambda_{\mathrm{W}}\langle\rho_{\mathrm{W}}(x,t)\rangle_i. \label{eq:rhotot}
\end{equation}
Starting from Eq.~(\ref{eq:deltarho}) we now proceed by evaluating the Wigner contribution in Eq.~(\ref{eq:rhotot}) for both a sudden and an adiabatic quench. In the former case one obtains
\begin{equation}
\langle\rho_{\mathrm{W}}^{\mathrm{sq}}(x,t)\rangle_i\!\approx\!\frac{N_\rho}{L}\frac{\pi N_\rho}{K_\rho}\!\int_{0}^{t}\!\frac{dt'}{\tau_0}\!\sum_{\eta=\pm}\!\eta\sin\left[\mathcal{L}_2\left(N_\rho,\mathcal{S}^{\mathrm{sq}}_{\eta}(x,t,t')\right)\right]\!\!\left[E_\rho^{\mathrm{sq}}\left(\mathcal{S}^{\mathrm{sq}}_{\eta}(x,t,t'),t'\right)\right]^{2},\label{eq:rhoWsq}
\end{equation}
where $ \mathcal{S}^{\mathrm{sq}}_{\pm}(x,t,t')=x \pm v_\rho(t'-t) $ and $\tau_0=L/v_F$ is the typical time-scale of the system.
Following a similar procedure for the two regions of the adiabatic quench case we obtain
\begin{eqnarray}
\langle\rho_{\mathrm{W}}^{\mathrm{II,ad}}(x,t)\rangle_i,&\approx&\frac{N_\rho}{L}\frac{\pi N_\rho}{K_\rho}\int_{0}^{t}\frac{dt'}{\tau}\sum_{\eta=\pm}\Big\{\eta\sin\left[\mathcal{L}_2\left(N_\rho,\mathcal{S}^{\mathrm{II,ad}}_{\eta}(x,t,t')\right)\right]\nonumber\\
&\times&\left[E_\rho^{\mathrm{II,ad}}\left(\mathcal{S}^{\mathrm{II,ad}}_{\eta}(x,t,t'),t'\right)\right]^{2}\Big\},\label{eq:rhoWadII}
\end{eqnarray}
with $ \mathcal{S}^{\mathrm{II,ad}}_\pm(x,t,t')=x\pm[\ell_\rho(t')-\ell_\rho(t)] $ and 
\begin{eqnarray}
\langle\rho_{\mathrm{W}}^{\mathrm{III,ad}}(x,t)\rangle_i&\approx&\frac{N_\rho}{L}\frac{\pi N_\rho}{K_\rho}\bigg\{\int_{0}^{\tau}\frac{dt'}{\tau}\sum_{\eta=\pm}\eta\sin\left[\mathcal{L}_2\left(N_\rho,\mathcal{S}^{\mathrm{m,ad}}_{\eta}(x,t,t')\right)\right]\nonumber\\
&\times&\left[E_\rho^{\mathrm{II,ad}}\left(\mathcal{S}^{\mathrm{m,ad}}_{\eta}(x,t,t'),t'\right)\right]^{2}+\int_{\tau}^{t}\frac{dt'}{\tau_0}\sum_{\eta=\pm}\eta\sin\left[\mathcal{L}_2\left(N_\rho,\mathcal{S}^{\mathrm{III,ad}}_{\eta}(x,t,t')\right)\right]\nonumber\\
&\times&\left[E_\rho^{\mathrm{III,ad}}\left(\mathcal{S}^{\mathrm{III,ad}}_{\eta}(x,t,t'),t'\right)\right]^{2}\bigg\},\label{eq:rhoWadIII}
\end{eqnarray}
where $ \mathcal{S}_\pm^{\mathrm{m,ad}}(x,t,t')=x\pm[\ell_\rho(t')-\ell_\rho(\tau)-v_\rho (t-\tau)] $ and $ \mathcal{S}_\pm^{\mathrm{III,ad}}(x,t,t')=x\pm v_\rho(t'-t) $. 

\section{Results}\label{sec:results}
\begin{figure}[htbp]
	\begin{center}
		\includegraphics[height=0.24\textheight]{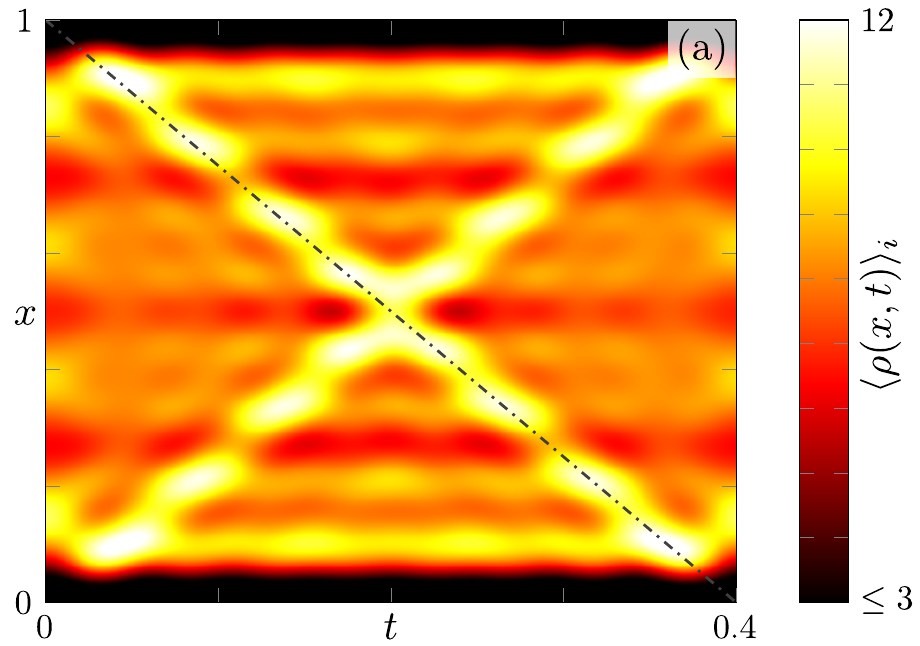}
		\includegraphics[height=0.24\textheight]{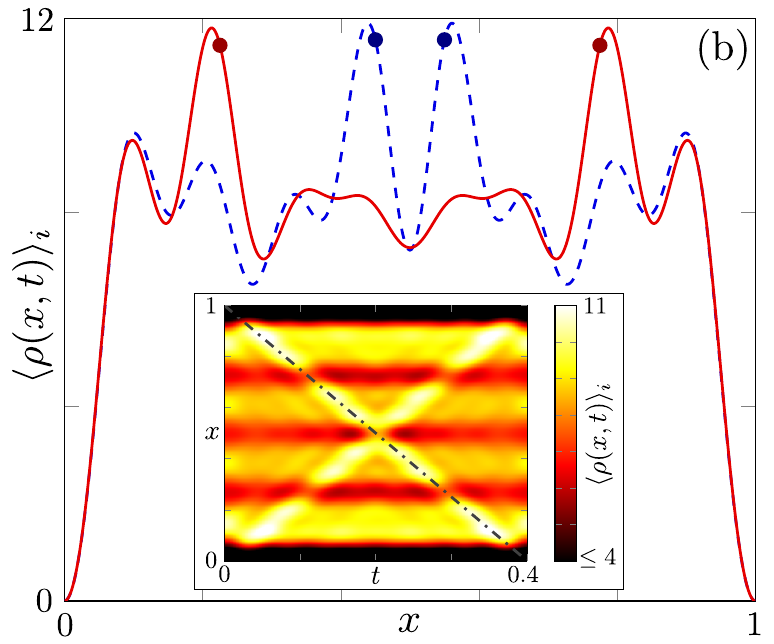}\\
		\caption{Panel (a): density plot of the total average density $ \langle\rho(x,t)\rangle_i $ as a function of $ x $ (units $ L $) and $ t $ (units $ \tau_0 $) over one period $ \mathcal{T}_\rho $ for a sudden quench from $ K_{\rho,i}=1 $ to $ K_\rho=0.4 $ with $ \lambda_{\mathrm{W}}=0.5 $. The dash-dotted gray lines highlight the position of one of the two LC perturbations. Panels (b): snapshot of the density for $ t^\star=0.09\tau_0 $ (red smooth line) and $t^\star=0.18\tau_0 $ (blue dashed line) for $ \lambda_{\mathrm{W}}=0.5 $. The thick dots represent the positions of the LC perturbations. Inset: same as Panel (a) for $ \lambda_\mathrm{W}=0.3 $. In all Panels, $ N_\uparrow=N_\downarrow=4 $.}
		\label{fig:rhosq}
	\end{center}
\end{figure}

In this Section we investigate the dynamics of the average density of Eq.~(\ref{eq:rhotot}) for both a sudden and an adiabatic switching on of the inter-particle interaction. The initial average density possesses only long-wave and Friedel contribution and thus, assuming $ N_\rho $ even, exhibits $ N_\rho/2 $ peaks~\cite{Fabrizio:1995,GindikinSablikov,Gambetta:2014}. After turning on the umklapp Hamiltonian a Wigner contribution starts growing in the average density. Thus, if the final inter-particle interaction is strong enough, one expects the average density to exhibit at some point the $ N_\rho $ peaks associated to the Wigner term, signaling the formation of a Wigner molecule inside the system~\cite{wigner,GindikinSablikov,Soffing:2009,wignernoi}. As we will show below, the dynamics of this molecule is dramatically affected by the time duration of the quenching protocol and its emergence is conveyed by LC perturbations propagating through the system~\cite{lightcone1,Porta:2016}.
\paragraph{Sudden quench} Figure~\ref{fig:rhosq} shows the dynamics of $ \langle\rho(x,t)\rangle_i $ for a system with $ N_\uparrow=N_\downarrow=4 $ fermions after a sudden quench of the inter-particle interaction from the non-interacting case with $ K_{\rho,i}=1 $ to $ K_\rho=0.4 $ -- corresponding to an equilibrium situation with a well formed Wigner molecule -- for $ \lambda_\mathrm{W}=0.5 $.
From Panels (a) it can be clearly seen that the post-quench density dynamics exhibits a peculiar LC behavior. Indeed, at $ t=0 $ two LC perturbations arise from the edges of the system and start traveling ballistically in opposite directions with velocity $ \pm v_\rho $,  i.e. the velocity of the final bosonic excitations, bouncing elastically whenever they reach one of the system boundaries~\cite{Porta:2016}. The LC perturbations are encoded in the functional dependence $ x\pm v_\rho t $ of the envelope function in Eq.~(\ref{eq:envsq}), which appears both in Friedel and Wigner contributions~\cite{Porta:2016}. Furthermore, in the latter term they also emerge due to the presence of the function $ \mathcal{S}^\mathrm{sq}_\eta(x,t,t') $ in the integrand of  Eq.~(\ref{eq:rhoWsq}). Comparing Panel (a) with the inset in Panel (b), where the same quench of Fig.~\ref{fig:rhosq}(a) for $ \lambda_{\mathrm{W}}=0.3 $ is shown, one can see that this latter contribution enhances the overall amplitude of the LC perturbations. As a further effect, the quench of the umklapp perturbation also produces some ripples in the density -- see  Fig.~\ref{fig:rhosq}(a) -- which, however, do not produce significant modifications in the density. From Fig.~\ref{fig:rhosq} clearly emerge the relevance of the LC perturbations in the formation and evolution of the Wigner molecule inside the system: In regions not reached by any LC perturbations the density exhibits essentially the same behavior as in the initial state, oscillating with Friedel wave vector $ \pi N_\rho/L $, while where one of the two LC perturbations has passed the original Friedel peaks split and oscillations with Wigner wave vector $ 2\pi N_\rho/L $ emerge. However, the newly formed Wigner molecule is destroyed by the passing of the second LC perturbation, which contributes to the restoration of the initial pre-quench situation and ensures the periodicity of the system dynamics~\cite{Bocchieri:1957}.

\begin{figure}[htbp]
	\begin{center}
		\includegraphics[width=\columnwidth]{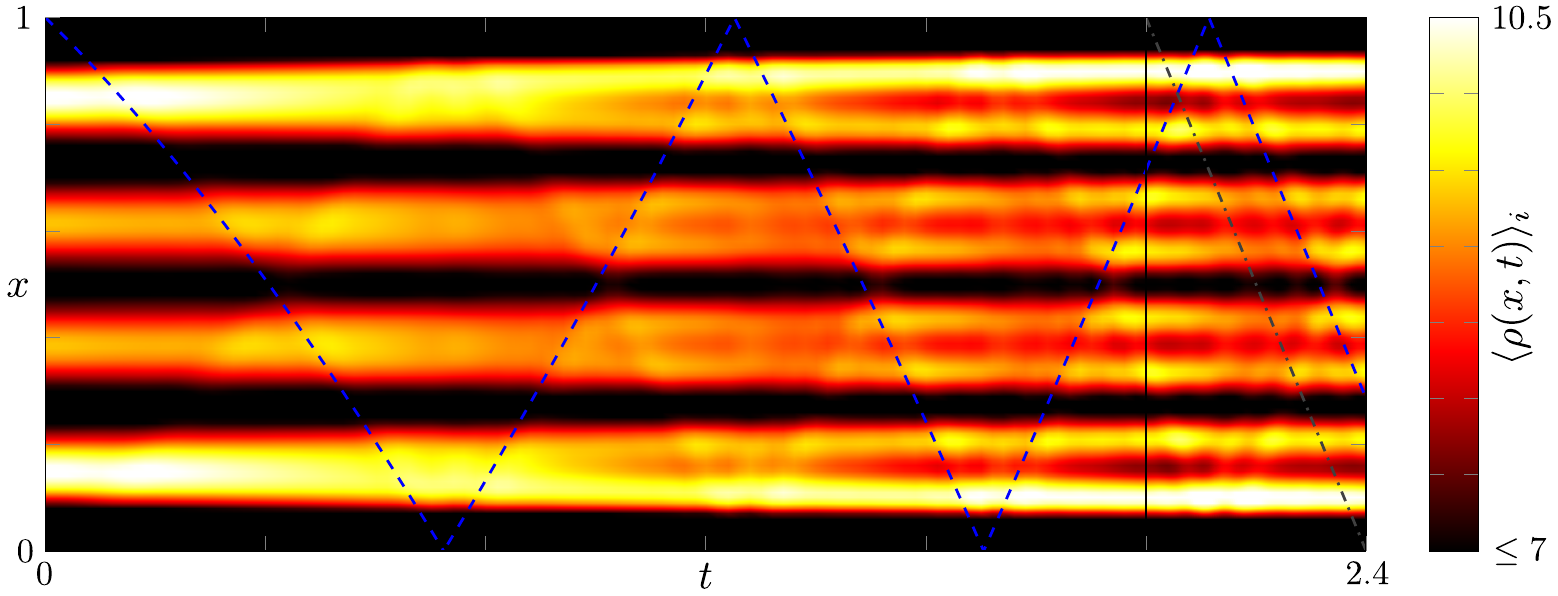}
		\caption{Density plot of the total average density $ \langle\rho(x,t)\rangle_i $ as a function of $ x $ (units $ L $) and $ t $ (units $ \tau_0 $) after an adiabatic quench of the inter-particle interaction from $ K_{\rho,i}=1 $ to $ K_\rho=0.4 $ with time duration $ \tau=2\tau_0 $. Here, two different counter-propagating  LC perturbations are present: the first ones (LC1) arise at $ t=0 $ (blue, dashed lines) while the second ones (LC2) emerge at $ t=\tau $ (gray, dash-dotted lines). For both LC1 and LC2 only one of the two LC perturbations is drawn. The black line represents $ t=\tau $. Here, $ N_\uparrow=N_\downarrow=4 $ and $ \lambda_\mathrm{W}=0.5 $. }
		\label{fig:rhoad}
	\end{center}
\end{figure}

\paragraph{Adiabatic quench} In order to create a stable Wigner molecule inside the system one has to resort to a slower quench. Figure~\ref{fig:rhoad} shows the behavior of $ \langle\rho(x,t)\rangle_i $ in a system with $ N_\uparrow=N_\downarrow=4 $ after an adiabatic quench of the inter-particle interaction from the non-interacting case with $ K_{\rho,i}=1 $ to $ K_\rho=0.4 $ with time duration $ \tau=2\tau_0 $ and $ \lambda_\mathrm{W}=0.5 $.
Let us analyze the dynamics of the system starting from $ t=0 $. Here, similarly to the sudden regime, two counter-propagating LC perturbations (LC1) arise from the system edges. In this case, however, they travel ballistically through the system with velocity $ \bar{v}_\rho(t) $, which corresponds to the {\it instantaneous} velocity of the LL bosonic modes [see Eq.~(\ref{eq:vinstgen})]~\cite{Bernier:2014,Porta:2016}. The presence of these perturbations is again encoded in the envelope function of Eq.~(\ref{eq:envadII}) and, in particular, in the function $ \mathcal{C}(x,\ell_\nu(t)) $~\cite{Porta:2016}. Moreover, as in the sudden quench case, the peculiar structure of the function $ \mathcal{S}^{\mathrm{II,ad}}_\eta(x,t,t') $ present in the Wigner contribution of Eq.~(\ref{eq:rhoWadII}) further enhances the LC amplitude. Even in this case the evolution of the density is strictly connected with the presence of LC1. Indeed, as can be seen from Fig.~\ref{fig:rhoad}, the density profile at position $ x $ remains approximately unchanged until it is reached by one of the two LC1 perturbations. However, in sharp contrast with the sudden case, after each subsequent passage of LC1 every pre-quench Friedel peak becomes broader and broader until it splits in two distinct Wigner peaks, signaling that a Wigner molecule begins to form inside the system. When the quench stops at $ t=\tau $, if the quench has lasted long enough as in the case of Fig.~\ref{fig:rhoad}, a Wigner molecule, created by the propagation of LC1 during the quench, has stabilized inside the system. In addition to this molecule one can see LC1 continuing its motion with velocity $ v_\rho $, as can be inferred from the second contribution to $ E^{\mathrm{III,ad}}_\nu(x,t) $ in Eq.~(\ref{eq:envadIII}) [see in particular the function $ \mathcal{C}(x,v_\nu(t-\tau)+\ell_\nu(\tau)) $] and from the functional form of $ \mathcal{S}^{\mathrm{m,ad}}_\pm(x,t,t') $ in Eq.~(\ref{eq:rhoWadIII})\footnote{Here, in the LC interpretation $ \ell(\tau) $ represents the distance modulo $ L $ traveled by LC1 during the ramp duration.}. However, at $ t=\tau$ two new counter-propagating LC perturbations (LC2) emerge from the system boundaries and start propagating ballistically with velocity $ v_\rho $. Again, these perturbations are encoded both in the third contribution to $ E^{\mathrm{III,ad}}_\nu(x,t) $ in Eq.~(\ref{eq:envadIII}) [see the function $ \mathcal{C}(x,v_\nu(t-\tau)) $] and in the form of $ \mathcal{S}^{\mathrm{III,ad}}_\pm(x,t,t') $ in Eq.~(\ref{eq:rhoWadIII}). Furthermore, from Eq.~(\ref{eq:envadIII}) and from Fig.~\ref{fig:rhoad} one can see that LC1 and LC2 have opposite effects, as in the spinless case~\cite{Porta:2016}: LC1 leads to an overall increasing of the density while LC2 tends to lower it. The post-quench dynamics is thus deeply affect by the interplay between LC1 and LC2 and this fact can be exploited to further increase the adiabaticity of the quench, as discussed in Ref.~\cite{Porta:2016} for the spinless case. Note that the more the quench is adiabatic, the less LC perturbations influence the post-quench dynamics of the Wigner molecule [see in particular Eq.~(\ref{eq:envadIII})], which will thus become more and more stable. 

\section{Conclusions}\label{sec:conclusions}
In this paper we investigated the density dynamics of a spinful LL with open boundary conditions after a sudden and an adiabatic quench of the inter-particle interaction. As a result of the latter, we also considered the switching on of an umklapp term in the system Hamiltonian, showing that it further enhances the LC effects arising due to the inter-particle interaction quench. We then studied the emergence of a Wigner molecule inside the system in both the quench regimes, showing that its formation is conveyed by LC perturbations arising at the beginning of the quench from the boundaries of the system. In particular, in the sudden quench case we showed that these perturbations creates an unstable Wigner molecule, with the density oscillating between this correlated state and the initial non-interacting one. On the other hand, in the adiabatic regime we obtained that during the quenching protocol each passing of the LC perturbations broaden a bit the initial Friedel peaks until they are split in two well separated Wigner peaks. When the quench stops two new LC perturbations arise from the boundaries and, together with the continuation of the initial ones, propagates through the system on top of the approximately stable Wigner molecule formed during the quench. 

\acknowledgments
We wish to thank F. Cavaliere, N. Traverso Ziani and M. Sassetti for useful discussions and the Italian Physical Society (SIF) for giving us the opportunity to realize this paper.

\end{document}